\documentstyle{article}
\title{
\bf Ising (anti-)ferromagnet on\\
    dynamical triangulations \\
    and quadrangulations.
}
\author{ D.A. Johnston \\
         Dept. of Mathematics\\
         Heriot-Watt University\\
         Riccarton\\
         Edinburgh, EH14 4AS\\
         Scotland}
\date { 18 June 1993 }
\begin{document}
  \maketitle
                      {\Large
                      \begin{abstract}
%
We write down matrix models for Ising spins with zero external field on
the vertices of dynamical triangulated random surfaces (DTRS) and
dynamically quadrangulated random surfaces (DQRS) and compare these
with the standard matrix model approach which places the spins on the
dual $\phi^3$ and $\phi^4$ graphs.  We show that the critical
temperatures calculated in the DTRS and DQRS models agree with those
deduced from duality arguments in the standard approach.  Using the
DQRS model we observe that the Ising antiferromagnet still undergoes a
phase transition to a Neel (checkerboard) ordered ground state which is
absent because of frustration in the other cases.  \\ \\To appear in
Physics Letters B.
%
			\end{abstract} }
%
  \thispagestyle{empty}
%
%
  \newpage
%
		  \pagenumbering{arabic}
The work of Boulatov and Kazakov \cite{1} solving the Ising model on
dynamical surfaces predated both the recent explosion of interest in
matrix models \cite{2} and the use of conformal field theory methods
for $c<1$ models coupled to 2d gravity \cite{3}. They used techniques
developed in earlier work by Brezin et.al. and Mehta \cite{4} to cast
the Ising model on dynamical $\phi^4$ and $\phi^3$ graphs (ie the duals
of DQRS and DTRS respectively) as two matrix integrals. The partition
function of the Ising model on some set of random graphs $G^n$ with $n$
vertices can be written as \begin{equation} Z_n(\beta, H) = \sum_{G^n}
\sum_{\{ \sigma \}} \exp \beta \left( \sum_{<ij>} G^n_{ij} \sigma_i
\sigma_j + H \sum_i \sigma_i \right) \label{e01} \end{equation} where
$G^n_{ij}$ is the connectivity matrix for a given graph.  Summing over
the number of vertices gives \begin{equation} Z(c, g, H) =
\sum_{n=1}^{\infty} \left( { - 4 g c \over (1 - c^2)^2 } \right)^n Z_n
(\beta, H) \label{e02} \end{equation} where $c = \exp ( - 2 \beta)$.
Boulatov and Kazakov showed that $Z$ was given by the free energy of
two $N \times N$ hermitian matrices $U$ and $V$ which in the case of
$\phi^4$ graphs with external field $H=0$ was:  \begin{equation}
F_{\phi^4} (c, g, H) = {1 \over N^2} \log \left( \int d^{N^2} M_1
d^{N^2} M_2 \exp  - tr {N \over g} \left( {M_1^2 \over 2} + {M_2^2
\over 2} - c M_1 M_2 + { 1 \over 4} M_1^4 + {1 \over 4} M_2^4 \right)
\right) \label{e03} \end{equation} and for $\phi^3$ graphs:
\begin{equation} F_{\phi^3} (c, g, H) = {1 \over N^2} \log \left( \int
d^{N^2} M_1 d^{N^2} M_2 \exp  - tr {N \over g} \left( {M_1^2 \over 2} +
{M_2^2 \over 2} - c M_1 M_2 + { 1 \over 3} M_1^3 + {1 \over 3} M_2^3
\right) \right).  \label{e03a} \end{equation} In these equations the
$M_1$ matrices can be thought of as representing $\sigma = +1$ vertices
and the $M_2$ matrices $\sigma = -1$ vertices. Rescaling to obtain
$\phi^4$ and $\phi^3$ couplings of $g/N$ and $g/\sqrt{N}$ respectively
shows that the correct topological factor of $V-E+F$ is obtained for
each graph, where $V$ is the number of vertices $E$ is the number of
edges (propagators) and $F$ is the number of faces.  Both the $\phi^4$
and $\phi^3$ models displayed a third order phase transition with
critical $c$ values of $c_{crit}^{\phi^4} = 1/4$ and $c_{crit}^{\phi^3}
= (\sqrt{28} - 1)/27$. The critical exponents were the same for both
lattices as one might have expected from universality and satisfied the
hyperscaling relations, but were different from the fixed lattice
Onsager exponents.  The values of these new exponents for the dynamical
case were confirmed by the Liouville theory calculations of KPZ and DDK
in \cite{3} in the light-cone and conformal gauges respectively.

The same universality that gives identical exponents on $\phi^{3}$ and
$\phi^{4}$ lattices might be expected to carry over to DTRS and DQRS
\footnote{This is not entirely trivial as it appears that situating
spins in the centre of simplices in the 3d equivalent gives very
different results from situating spins on the vertices \cite{6}} and
numerical simulations \cite{5} provide some confirmation of this,
giving satisfactory agreement with the predictions of \cite{1,3} for
both DTRS and $\phi^3$ graphs.  It would be reassuring, however, to
write down directly matrix models for Ising spins on DTRS and DQRS
instead of relying on duality arguments to find the critical
temperature and universality to find the exponents.  We have another
motivation for considering such models:  An Ising {\it
anti-}ferromagnet will display a phase transition to a Neel ordered
ground state on ``loosely packed'' lattices where only even loops are
present \cite{5a}. This is clear from the strong coupling expansion of
equ.(\ref{e01}) \begin{equation} Z_{V}(\beta, 0) = \sum_{G^V}  2^{V}
(\cosh ( \beta) )^{E} \sum_{loops} (\tanh (\beta))^{length} \label{e04}
\end{equation} where the sum runs over closed loops with each edge
traversed only once. Equ.(\ref{e04}) possesses a $\beta \rightarrow  -
\beta$ symmetry if all possible loops are even in length, which
translates into a mapping between the ferromagnetic ($\beta$ positive)
and antiferromagnetic ($\beta$ negative) partition functions and hence
a phase transition in the antiferromagnet. The $\phi^4$ and $\phi^3$
Ising models will contain loops of both even and odd length so
frustration will suppress the antiferromagnetic transition. The $\beta
\rightarrow - \beta$ symmetry which corresponds to $c \rightarrow 1/c$
should hold at the level of the partition function or free energy and
it is clear from equ.(\ref{e03}) and equ.(\ref{e03a}) that the $\phi^4$
and $\phi^3$ matrix models do not possess this symmetry. An
antiferromagnetic transition will also be absent on DTRS, but for DQRS
we would expect to see a transition as all loops must be of even
length. The matrix model for Ising spins on DQRS should therefore
possess a $\beta \rightarrow - \beta$ symmetry.

To write down the appropriate matrix models we change our perspective
slightly and think of the potential terms in the matrix integral as
representing the {\it faces} of the graph rather than the vertices
\cite{7}. For a one-matrix integral with action of the form $tr(
\phi^2/2 + ( g/\sqrt{N}) \phi^3)$ this still gives the correct counting
as each internal vertex gives a factor of N and each face (triangle in
this case) gives a factor of $N^{-1/2}$ to give a total of $N^{V -
F/2}$. As $3F = 2E$ this produces the correct factor of $N^{\chi}$
\footnote{I would like to thank Jan Ambj\o rn for sorting out my
confusion on this point}. Similarly, an action of the form $tr(
\phi^2/2 + ( g/N ) \phi^4)$ which generates squares will give $N^{V -
F}$ and in this case $4F = 2E$, so we again find $N^{\chi}$.  Let us
now apply similar considerations to the Ising model on DTRS with the
external field $H$ set to zero for simplicity. The triangular faces
will be composed of two types of edges, $S$ with the spins at both ends
the same, and $D$ with the spins at both ends different. Each $S$ edge
will contribute a factor \begin{equation} \cosh (\beta) + \sinh
(\beta)  = \cosh (\beta) ( 1 + c^*) \label{e05} \end{equation} and each
$D$ edge will contribute a factor \begin{equation} \cosh (\beta) -
\sinh (\beta)  = \cosh (\beta) ( 1 - c^*) \label{e06} \end{equation}
where we have introduced $c^* = (1 - c) / (1 + c)$, which is the
transformation effected on $c$ by taking the dual transformation of the
temperature $\beta^* = - 1/2 \log \tanh (\beta)$.  There will be two
possible types of triangles in the model, one with all the spins the
same giving a $tr S^3$ term in the action, and one with two spins the
same and one different giving an $tr SD^2$ term. We choose to put the
factors in equs.(\ref{e05},\ref{e06}) in the propagators so the matrix
model action we arrive at is \begin{equation} U_{DTRS} =  { N \over g}
tr \left( {1 \over 2 \cosh (\beta) (1 + c^*)} S^2 + { 1 \over 2 \cosh (
\beta) ( 1 - c^*)} D^2 + S^3 / 3 + S D^2 \right) \label{e07}
\end{equation} where $S$ and $D$ are again $N \times N$ hermitian
matrices.

We can show that this is an Ising model with a critical temperature
agreeing with that deduced from duality by transforming it to an $O(1)$
representation of the Ising model derived by Eynard and Zinn-Justin
\cite{8}.  A rescaling $S \rightarrow S / \cosh (\beta)$, $D
\rightarrow \sqrt{2 (1-c^*)}D / \cosh ( \beta)$, $g \cosh^3 (\beta)
\rightarrow g$ gives \begin{equation} U_{DTRS} =  { N \over g} tr
\left(  D^2 ( 1 + 2 S ( 1 - c^*) ) +  {1 \over 2  (1 + c^*)} S^2 + S^3
/ 3  \right).  \label{e08} \end{equation} The first term can be
simplified at the expense of introducing a more complicated set of
interactions by defining $S_1 = 1 + 2 S ( 1 - c^*)$ which gives, on
scaling $S_1 \rightarrow S_1/ ( 1 + c^*)$, $2 g ( 1 - c^*)^3 ( 1 +
c^*)^3 \rightarrow g$, $2 D^2 (1 -c^*)^3  (1+c^*)^2 \rightarrow D^2$
and dropping constant terms \begin{equation} U_{DTRS} = { N \over g} tr
\left( D^2 S_1 + { S_1^3 \over 12} - { c^*  \over 2 } S_1^2 + { (3c^* -
1)
 ( 1 + c^*) \over 4}  S_1 \right).  \label{e09} \end{equation}

We have carried out these manipulations because it was shown in
\cite{8} that defining $A = ( M_1 - M_2)/2$, $S = M_2 +  M_2 + 1 + c$
transformed the $\phi^3$ action in equ.(\ref{e03a}) into
\begin{equation} U_{\phi^3} = { N \over g} tr \left( A^2 S + { S^3
\over 12} - {c \over 2} S^2 + { (3 c - 1) (1 + c)\over 4} S \right)
\label{e10} \end{equation} which is identical in form to
equ.(\ref{e09}). Performing a gaussian integration over $A$ in this
action allowed saddle point methods to be used to determine the
critical behaviour of the model in the large $N$ limit. It was shown
that the resolvent of the matrix $S$ \begin{equation} \omega_0 (z) = {1
\over N} tr {1 \over z - S} \label{e11} \end{equation} was given by
\begin{equation} \omega_0 (z) = {1 \over 3 g} ( 2 V'(z) - V' ( -z)) + {
1 \over 12 g } \omega ( z ) \label{e12} \end{equation} where we have
written the action as $A^2 S +  V ( S)$ and $\omega ( z ) = e^{- 2 \pi
i \over 3} \omega_+ ( z ) + e^{ 2 \pi i \over 3} \omega_- ( z )$ with,
for a cubic potential $V(z)$ \begin{equation} \omega_{\pm} (z) = \left(
\sqrt{ 1 - b^2 / z^2} \mp i b / z \right)^{1/3} \left( \sqrt{ 1 - b^2 /
z^2} \pm i b \sigma / z\right).  \label{e13} \end{equation} For a
potential of the form  $4 V ( S ) = S^3 / 3 - \alpha S^2 + \beta S$ the
parameters in the solution $b$ and $\sigma$ were related to the
parameters in the potential $\alpha$ and $\beta$ by \begin{eqnarray} b(
3 \sigma - 1) &=& 6 \sqrt{3} \alpha \nonumber \\ b^2 ( 3 \sigma - 5 )
&=& 9 \beta.  \label{e14} \end{eqnarray} The critical Ising model
corresponded to $\sigma = 1$ which gave, on eliminating $b$,
\begin{equation} 6 \alpha^2 + \beta^2 = 0.  \label{e15} \end{equation}
If we substitute the values for the $\phi^3$ model ($\alpha = 2 c$,
$\beta = (3c - 1) ( 1 + c)$) into equ.(\ref{e15}) we get $27c^2 + 2 c -
1 = 0$, which reproduces the correct value of $c_{crit}^{\phi^3} =
(\sqrt{28} - 1) / 27$.  For the DTRS model we replace $c$ by $c^*$
which on substituting into equ.(\ref{e15}) gives $27(c^*)^2 + 2 c^* - 1
= 0$. The critical value of $c^*$ on DTRS is thus $(\sqrt{28} - 1) /
27$ so the critical value of $c$ ($= ( 1 - c^*)/(1 + c^*)$) is
$c^{DTRS}_{crit} = (14 - \sqrt{7}) / (13 + \sqrt{7})$, which is the
dual of the value on $\phi^3$ graphs.  It is also worth remarking that
$\beta \rightarrow - \beta$ transforms $c^* \rightarrow - c^*$ which is
obviously not a symmetry of equ.(\ref{e09}), so there will be no
antiferromagnetic transition on DTRS.

To write down the Ising model on DQRS we use similar notation,
representing same spin edges by $S$ and different spin edges by $D$. In
this case we will attach the weights to the interaction terms that
generate the various possible square faces, though it is also possible
to attach the inverse factors to the propagators.  The action is given
by \begin{equation} U_{DQRS} = {N \over g} tr \left( {1 \over 2} S^2  +
{ 1 \over 2} D^2 + {g_1 \over 4} S^4 + {g_2 \over 4} D^4 + {g_3 \over
2} ( S D S D + 2 S^2 D^2 ) \right) \label{e16} \end{equation} where
\begin{eqnarray} g_1 &=& \cosh^2 ( \beta) ( 1 + c^*)^2 \nonumber \\ g_2
&=& \cosh^2 ( \beta) ( 1 - c^*)^2 \nonumber \\ g_1 &=& \cosh^2 ( \beta)
( 1 + c^*) (1 - c^*).  \label{e17} \end{eqnarray} These couplings are
the square root of those one might naively write, but it should be
remembered that each edge is shared by two squares. Using
\begin{equation} \cosh ( \beta ) ( 1 \pm c^* ) = { 1 \over \cosh (
\beta ) ( 1 \mp c^* )} \label{e18} \end{equation} and scaling $S
\rightarrow S  \cosh ( \beta )$, $D \rightarrow D  \cosh ( \beta )$, $g
\rightarrow g \cosh^2 ( \beta )$ we can write this as \begin{equation}
U_{DQRS} = {N \over g} tr \left( {1 \over 2} S^2  + { 1 \over 2} D^2 +
{1 \over 4 ( 1 - c^*)^2} S^4 + {1\over 4 ( 1 + c^*)^2} D^4
 + {1 \over 2 ( 1 - c^*) ( 1 + c^*)} ( S D S D + 2 S^2 D^2 ) \right).
\label{e19} \end{equation} We can massage the $\phi^4$ model of
equ.(\ref{e03}) into a similar form by defining $A = ( M_1 + M_2 ) /
\sqrt{2}$, $B = (M_1 - M_2) / \sqrt{2}$ and then rescaling $A
\rightarrow \sqrt{2} A / \sqrt{ 1 - c}$, $B \rightarrow \sqrt{2} B /
\sqrt{ 1 + c}$, $g \rightarrow 2 g $ to get \begin{equation} U_{\phi^4}
= {N \over g} tr \left( {1 \over 2} A^2  + { 1 \over 2} B^2 + {1 \over
4 ( 1 - c)^2} A^4 + {1\over 4 ( 1 + c)^2} B^4
 + {1 \over 2 ( 1 - c) ( 1 + c)} ( A B A B + 2 A^2 B^2 ) \right).
\label{e20} \end{equation} which is identical to equ.(\ref{e19}) up to
the replacement of $c^*$ by $c$.  We know that $U_{\phi^4}$ in
equ.(\ref{e20}) represents an Ising model with $c^{\phi^4}_{crit} =
1/4$, so $U_{DQRS}$ in equ.(\ref{e19}) must represent an Ising model
with $c^*_{crit} = 1/4$. This gives $c^{DQRS}_{crit} = (1 -c^*_{crit})/
(1 + c^*_{crit}) = 3/5$, which is identical to the value deduced from
duality and the $\phi^4$ model result. In addition we can see that
$U_{DQRS}$ possesses an obvious $c^* \rightarrow - c^*$ symmetry if one
also exchanges $S$ and $D$ so we can deduce that the Ising
antiferromagnet on DQRS will display a phase transition.  One might
worry that the summation over quadrangulations would disrupt the Neel
order of the low temperature phase, casting doubt on the above
observation. We can see, however, that this is not so by considering
the ``flip'' moves that implement the sum over quadrangulations in
simulations.  The two possible flips for squares maintain a connection
between unlike spins and thus cost no energy in a Neel ground state -
see Fig.1.

To summarize, we have written down matrix models that represent Ising
spins on the vertices of DTRS and DQRS and shown that their critical
points agree with those deduced from duality for Ising spins on
$\phi^3$ and $\phi^4$ graphs.  Ising spins on DQRS also manifest an
antiferromagnetic transition because only even loops are possible in
this case.  It would be interesting to include a non-zero external
field in the models in this paper.  It is perhaps worth remarking in
closing that complex matrix models generate surfaces composed of only
even loops \cite{10} so one might also consider using a two complex
matrix model to examine the behaviour of the Ising antiferromagnet.

\bigskip \centerline{\bf Figure Caption} \begin{description} \item{Fig.
1.} The two possible flip moves on adjacent squares in DQRS both
preserve the ``checkerboard'' order of the ground state. Up spins are
shown as solid dots and down spins as open dots.  \end{description}
\vfill \eject
\end{document}